\documentclass[12pt,a4paper]{article}
\usepackage{amsmath}
\usepackage{amssymb}
\usepackage{bm}
\usepackage[dvips]{graphicx}
\usepackage{color}
\usepackage{url}
\usepackage{hyperref}

\catcode `@=11 \@addtoreset{equation}{section} \catcode `@=12

\begin{document}
\setlength{\oddsidemargin}{0cm}
\setlength{\baselineskip}{7mm}

\begin{titlepage}

	\begin{center}
		{\LARGE
		Application of Bootstrap to $\theta$-term
		}
	\end{center}
	\vspace{0.2cm}
	\baselineskip 18pt 
	\renewcommand{\thefootnote}{\fnsymbol{footnote}}

	\begin{center}

		Yu {\sc Aikawa}$^{a}$\footnote{%
		E-mail address: aikawa.yu.17(at)shizuoka.ac.jp}, 
		Takeshi {\sc Morita}$^{a,b}$\footnote{%
			E-mail address: morita.takeshi(at)shizuoka.ac.jp
		} and
		Kota {\sc Yoshimura}$^{c}$\footnote{%
		E-mail address: kyoshimu(at)nd.edu
	}

		\renewcommand{\thefootnote}{\arabic{footnote}}
		\setcounter{footnote}{0}
		
		\vspace{0.4cm}
		
		{\it
			a. Department of Physics,
			Shizuoka University \\
			836 Ohya, Suruga-ku, Shizuoka 422-8529, Japan 
			\vspace{0.2cm}
			\\
			b. Graduate School of Science and Technology, Shizuoka University\\
			836 Ohya, Suruga-ku, Shizuoka 422-8529, Japan
			\vspace{0.2cm}
			\\
			c. Department of Physics, University of Notre Dame \\
			Notre Dame, Indiana, 46556, USA
			}

	\end{center}
	
	
	\vspace{1.5cm}
	
	\begin{abstract}

	\end{abstract}
	Recently, novel numerical computation on quantum mechanics by using a bootstrap method was proposed by Han, Hartnoll, and Kruthoff.
    We consider whether this method works in systems with a $\theta$-term, where the standard Monte-Carlo computation may fail due to the sign problem.
    As a starting point, we study quantum mechanics of a charged particle on a circle in which a constant gauge potential is a counterpart of a $\theta$-term.
    We find that it is hard to determine physical quantities as functions of $\theta$ such as $E(\theta)$, except at $\theta=0$ and $\pi$.
    On the other hand, the correlations among observables for energy eigenstates are correctly reproduced for any $\theta$.
    Our results suggest that the bootstrap method may work not perfectly but sufficiently well, even if a $\theta$-term exists in the system.    
	
\end{titlepage}

\section{Introduction}

Numerical analysis plays essential role in modern physics.
Without the help of numerical analysis, quantitative evaluation is very difficult in many cases.
Particularly, Monte-Carlo computation (MC) is quite powerful and is widely employed in various studies.
However, MC may not be available when there are sign problems.
Hence, various alternative numerical approaches such as complex Langevin method \cite{PhysRevA.29.2036, Parisi:1983mgm, PhysRevD.81.054508,Aarts:2011ax, Nagata:2016alq, Nagata:2016vkn}, tensor renormalization method \cite{Levin:2006jai} and Lefschetz thimble method \cite{Witten:2010cx, PhysRevD.86.074506} are being studied actively.
(See a review article \cite{Nagata:2021bru} for recent progress on lattice QCD.)

Recently, as a novel numerical tool, the bootstrap method was proposed \cite{Anderson:2016rcw, Lin:2020mme} in zero-dimensional matrix models.
This method was applied to quantum mechanics  \cite{Han:2020bkb}, and several studies confirmed its validity \cite{Kazakov:2021lel, Berenstein:2021dyf, Bhattacharya:2021btd}.
Particularly, this method works in gauge theories at large-$N$ limit \cite{Anderson:2016rcw, Lin:2020mme, Han:2020bkb, Kazakov:2021lel, Bhattacharya:2021btd}. (Indeed, finite $N$ is harder in this method.)
Since taking large-$N$ limit is difficult in MC, 
the bootstrap method may provide a new window of numerical study of large-$N$ gauge theories.

Then, it is natural to ask whether the bootstrap method is applicable to systems with sign problems.
In this article, we study a $\theta$-term, which is pure imaginary in the Euclidean action and causes a sign problem.
Since the application of the bootstrap method to higher dimensional quantum field theories has not been established, 
as a starting point, we study quantum mechanics of a charged particle on a circle.
\begin{align}
    S(\theta)=& \int dt \left( \frac{1}{2} \dot x^2 -V(x) \right)  - \frac{\theta}{2\pi}\int dt \dot x, \nonumber \\
 & V(x) = a(1-\cos ( x) ).
    \label{S}
 \end{align}
 Here, we impose a periodicity $x=x+2\pi$ and $a$ is a non-negative coupling constant.
 $\theta$ is a real constant parameter and can be regarded as a background constant gauge potential, which causes the Aharonov-Bohm effect.
 The last term becomes a pure imaginary in the Euclidean action, and it is a counterpart of the $\theta$-term in QCD.
 Although this model is very simple, several similarities between this model and the $\theta$-term in QCD are known \cite{Tong:gauge}, and
 this model may provide us an intuition whether the bootstrap method potentially works in QCD or not.

Note that the numerical bootstrap method \cite{Han:2020bkb} employs the Hamiltonian formalism rather than the path integral formalism that uses the Euclidean action, and we may naively expect that the sign problem can be evaded in the bootstrap method.
However, we find that, although we can avoid the sign problem, the bootstrap method encounters another problem on a gauge fixing.
Due to this new issue, it is difficult to determine physical quantities as functions of $\theta$ such as $E(\theta)$ in the bootstrap method.
On the other hand, the correlations among observables for energy eigenstates are correctly obtained for any $\theta$.
For example, we can describe the expectation value of a position operator $\langle E | e^{ix} |E \rangle $ as a function of energy eigenvalues.
In addition, $\theta=0$ and $\pi$ are special, and physical quantities can be determined there.
These results suggest that the bootstrap method works not perfectly but sufficiently well, even if a $\theta$-term exists in systems.

The organization of this article is as follows.
In section \ref{sec-analytic}, we review the model \eqref{S} and derive the spectra and several quantities.
In section \ref{sec-bootstrap}, we employ the numerical bootstrap method and compare the obtained results with the ones derived in Sec.~\ref{sec-analytic}.
We also discuss the gauge fixing problem.
Section \ref{sec-discussion} contains conclusions and discussions.
In Appendix \ref{app-a=0}, we study the model at $a=0$ to get a insight of the system.
In Appendix \ref{app-bootstrap}, we briefly explain our numerical analysis.

\paragraph{Note:}
After we submitted our manuscript to arxiv, related works  were done independently in Refs.~\cite{Berenstein:2021loy, Tchoumakov:2021mnh}.
Particularly, a prescription for the gauge fixing problem was proposed in Ref.~\cite{Tchoumakov:2021mnh}.

\section{Analytic Study of the Model}
\label{sec-analytic}

In this section, we study the model \eqref{S} analytically, and we will compare the results in this section with the bootstrap analysis in the next section. We will call the results in this section as ``analytic results" in order to distinguish the results obtained through the bootstrap method.

From \eqref{S}, we obtain the Hamiltonian
\begin{align}
	H(\theta):=\frac{1}{2}\left( p +\frac{\theta}{2\pi}  \right)^2 -a(\cos ( x) -1),
	\label{H}
\end{align}
and we investigate the Schr$\ddot{\rm o}$dinger equation
\begin{align}
	E \psi(x) = H(\theta) \psi(x),
	\label{Sch-eq}
\end{align}
where $\psi(x)$ is an energy eigenfunction and $E$ is its energy eigenvalue.

Importantly, this system has a gauge symmetry.
Indeed, by multiplying $e^{i\frac{ \lambda}{2\pi} x}$ from the left in \eqref{Sch-eq}, 
we obtain
\begin{align}
	E e^{i\frac{ \lambda}{2\pi} x}\psi(x)= e^{i\frac{ \lambda}{2\pi} x} H(\theta) e^{-i\frac{ \lambda}{2\pi} x}e^{i\frac{ \lambda}{2\pi}x} \psi(x) 
	= H(\theta-\lambda) e^{i\frac{ \lambda}{2\pi}x} \psi(x), 
\end{align}
where $\lambda$ is a real constant parameter and we have used $e^{i\frac{ \lambda}{2\pi} x} p e^{-i\frac{ \lambda}{2\pi} x}=p-\lambda/2\pi$.
Thus, $\theta$ and the wave function are transformed as
\begin{align}
\theta \to \theta - \lambda, \qquad \psi(x) \to e^{i \frac{ \lambda}{2\pi} x} \psi(x) .
\label{gauge-tr}
\end{align}
Particularly, this transformation changes the periodicity of the wave function by $\lambda$.
For example, if the original wave function is periodic, the periodicity becomes
\begin{align}
 \psi(x+2\pi)= \psi(x) \to \tilde{\psi}(x) :=  e^{i\frac{ \lambda}{2\pi} x} \psi(x), \quad
 \tilde{\psi}(x+2\pi)=  e^{i\lambda } \tilde{\psi}(x).
\end{align}
Correspondingly, the momentum  $\langle p \rangle $ is transformed as
\begin{align}
	\langle p \rangle \to 	\langle p \rangle + \frac{1}{2\pi} \lambda .
	\label{gauge-tr-p}
\end{align}
Thus, the momentum $\langle p \rangle$ is not a gauge invariant observable.
On the other hand, the velocity 
\begin{align}
	\langle	\dot x  \rangle:= \langle p \rangle  + \frac{\theta}{2\pi}  ,
	\label{velocity}
\end{align}
is gauge invariant.

\begin{figure}
	\begin{tabular}{ccc}
		\begin{minipage}{0.33\hsize}
			\begin{center}
				\includegraphics[scale=0.4]{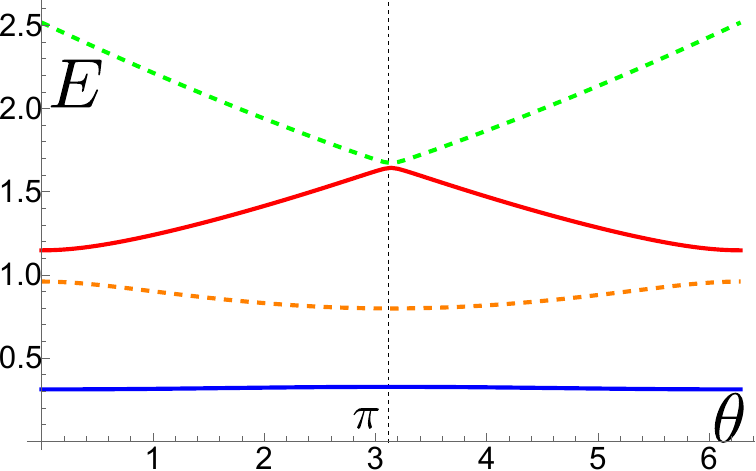}\\
				 $\theta$ vs. $E$
			\end{center}
		\end{minipage}		
		\begin{minipage}{0.33\hsize}
			\begin{center}
				\includegraphics[scale=0.4]{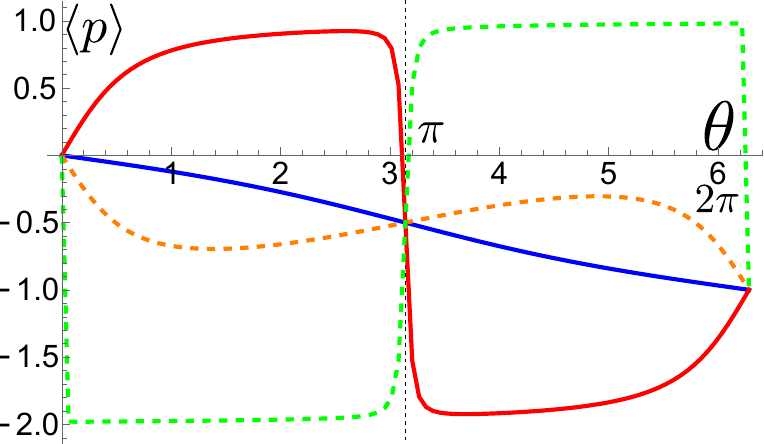}\\
				$\theta$ vs. $\langle p \rangle$
			\end{center}
		\end{minipage}
		\begin{minipage}{0.33\hsize}
			\begin{center}
				\includegraphics[scale=0.4]{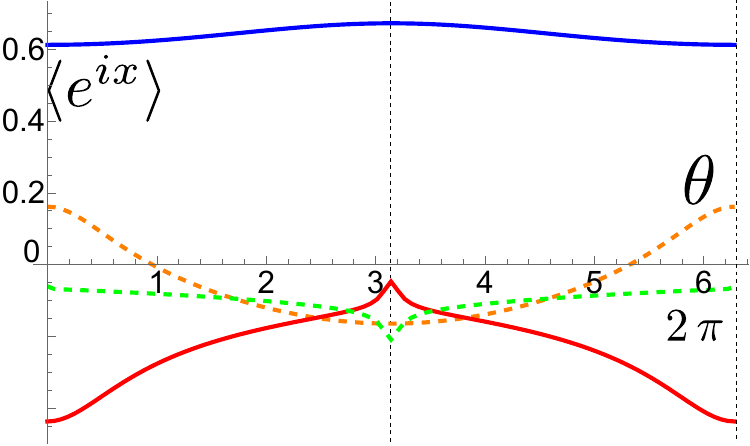}\\
				$\theta$ vs. $\langle e^{ix} \rangle$
			\end{center}
		\end{minipage}
    \end{tabular}
	\caption{$\theta$ dependence of $E$, $\langle p \rangle$ and   $\langle e^{ix} \rangle$.
    We numerically solve the Schr$\ddot{\rm o}$dinger equation \eqref{H} with the boundary condition $\psi(x+2\pi)=\psi(x)$.
    The curves are for the first four energy eigenstates: The blue curves, the orange dashed ones, the red ones and the green dashed ones are the first, the second, the third and the fourth state, respectively.
    We take $a=1/2$ in these plots.
		}
	\label{Fig-a1}
\end{figure}

Because of this gauge redundancy, we need to fix a gauge in order to solve the  Schr$\ddot{\rm o}$dinger equation.
In the following analysis, we fix the periodicity of the wave function periodic: $ \psi(x+2\pi)= \psi(x)$.
However, we can still apply a gauge transformation \eqref{gauge-tr} with $\lambda=2\pi $, which retains this periodicity, and $(\theta, \langle p \rangle)$ is identical to $(\theta-2\pi, \langle p \rangle+1 )$. 
Thus, the physical domain of $\theta$ can be taken as $0 \le \theta < 2\pi$.

Under this gauge fixing condition, we solve the Schr$\ddot{\rm o}$dinger equation \eqref{Sch-eq} and evaluate the energy spectrum, $\langle p \rangle$ and  $\langle e^{ix} \rangle$ for the energy eigenstates\footnote{
The Schr$\ddot{\rm o}$dinger equation \eqref{Sch-eq} may be solved in terms of Mathieu functions, but we use Mathematica package NDEigensystem.
}.
The results at $a=1/2$ are summarized in Fig.~\ref{Fig-a1}.
Here, we plot these quantities for the first four energy eigenstates with respect to $\theta$.

However, $\theta$ and $\langle p \rangle$  are not gauge invariant, and the correlations between the gauge invariant quantities may be more important.
Hence, we plot  $\langle \dot{x} \rangle$ vs. $E$ and $E$ vs. $\langle e^{ix} \rangle$ in Fig.~\ref{Fig-a1-obs}.
We will later see that the bootstrap method reproduces these correlations correctly.

In Appendix \ref{app-a=0}, some results at $a=0$ is shown.
There, the calculations are very simple, and they may provide intuitions about the properties of the solutions at $a>0$ plotted in Fig.~\ref{Fig-a1} and \ref{Fig-a1-obs}.

\begin{figure}
	\begin{tabular}{cc}
		\begin{minipage}{0.5\hsize}
			\begin{center}
				\includegraphics[scale=0.6]{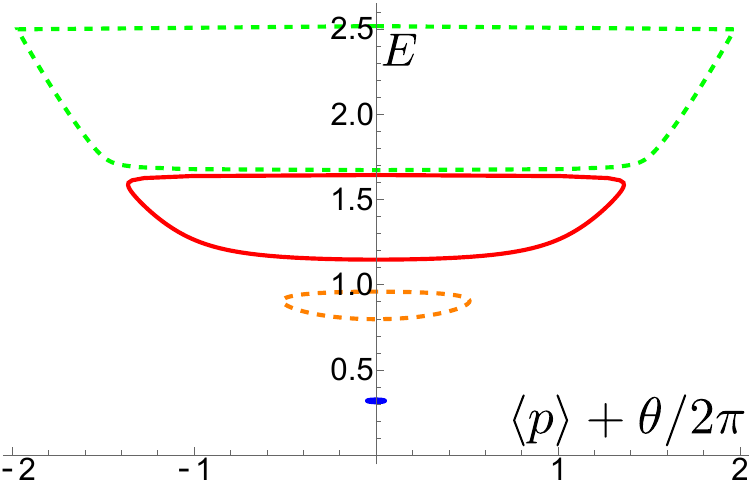}\\
				 $\langle \dot{x} \rangle=\langle p \rangle+\theta/2 \pi$ vs. $E$
			\end{center}
		\end{minipage}		
		\begin{minipage}{0.5\hsize}
			\begin{center}
				\includegraphics[scale=0.6]{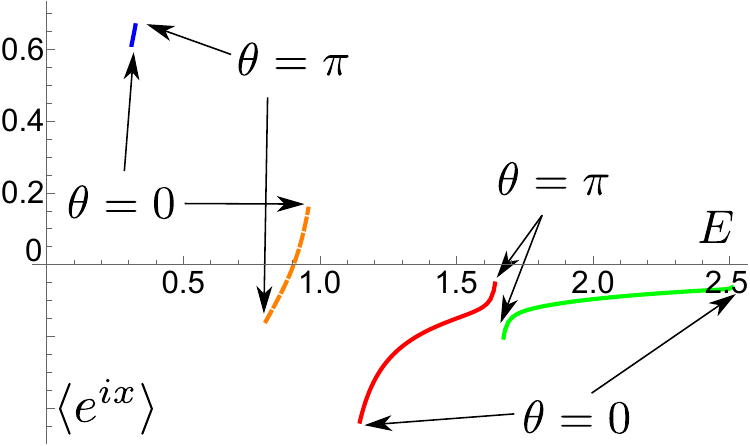}\\
				$E$ vs. $\langle e^{ix} \rangle$
			\end{center}
		\end{minipage}
	\end{tabular}
	\caption{Correlations among the gauge invariant quantities: $E$, $\langle \dot{x} \rangle$ and $\langle e^{ix} \rangle$.
    The curves are for the first four energy eigenstates considered in Fig.~\ref{Fig-a1}.
    The line styles correspond to those in Fig.~\ref{Fig-a1}.
		}
	\label{Fig-a1-obs}
\end{figure}

\section{Bootstrap Analysis}
\label{sec-bootstrap}

We analyze \eqref{H} via the bootstrap method and will compare the results with the analytic ones obtained in the previous section.
The details of our numerical computation is explained in Appendix \ref{app-bootstrap}.

First, we briefly introduce the numerical bootstrap method.
We consider the following operators\footnote{
	Note that we do not consider operators like $p^n  e^{i m  x}$ because they can be represented by the operators \eqref{observables} through the commutation relation and the final result would not change so much.
	Besides, we consider the operators $e^{  i mx}$ rather than $x^m$.
	This is because the operators $x^m$ are not well defined in quantum mechanics on $S^1$. 
	For example, it does not satisfy $\langle [H,x] \rangle=0$ even if the state is an energy eigenstate.
	We can easily confirm it in the $a=0$ case by using \eqref{a=0}.
	},
\begin{align}
O_{mn}:= e^{i m  x} p^n , \qquad m=0,\pm 1, \pm 2, \cdots, \quad n=0,1,2,\cdots. 
\label{observables}
\end{align}
Then, we define
\begin{align}
\tilde{O}= \sum_{m=0}^{K_x}  \sum_{n=0}^{K_p} c_{mn} O_{mn}
=c_{00} +  c_{10} e^{  i x}  +  c_{01} p + \cdots,
\label{operators-bootstrap}
\end{align}
where $\{ c_{nm} \}$ are constants, and $K_p$ and $K_x$ are non-negative integers.
Since $ \langle  \alpha | O^\dagger O | \alpha \rangle \ge 0$ is satisfied for any state  $  | \alpha \rangle $  in this system for arbitrary well-defined operators $O$, 
\begin{align}
\langle  \alpha | \tilde{O}^\dagger \tilde{O} | \alpha \rangle \ge 0
\end{align}
is satisfied for any constants $\{ c_{nm} \}$.
Hence, the following $(K_p+1)(K_x+1)\times (K_p+1)(K_x+1)$ matrix ${\mathcal M}$ has to be positive-semidefinite \cite{Han:2020bkb},
\begin{align}
{\mathcal M}:=
\begin{pmatrix}
1 & \left\langle \alpha | e^{  i x} | \alpha \right\rangle  & \left\langle  \alpha | p | \alpha \right\rangle   & \cdots \\
\left\langle \alpha | e^{-  i x} | \alpha \right\rangle & 1 & \left\langle \alpha | e^{-  i x} p | \alpha \right\rangle &  \cdots\\
\left\langle  \alpha | p | \alpha \right\rangle  &  \left\langle \alpha | p e^{  i x} | \alpha \right\rangle & \left\langle \alpha | p^2  | \alpha \right\rangle  & \cdots \\
\vdots & \vdots  & \vdots  & \ddots
\end{pmatrix}
\succeq 0,
\label{bootstrap}
\end{align}
This strongly constrains the possible values of the observables $\langle \alpha | O_{mn} | \alpha \rangle$.
We call ${\mathcal M}$ as a bootstrap matrix.
Note that, as $K_x$ and $K_p$ increase, the constraint becomes typically stronger.
 
From now on, we focus on energy eigenstates and take $|\alpha \rangle $ as an energy eigenstate  $|E \rangle $.
Then, $\langle E| O_{mn} |E \rangle$ has to satisfy the following two conditions \cite{Han:2020bkb}:
\begin{align}
&\langle E| \left[ H, O_{mn} \right] |E \rangle =0, 
\label{HO=0}
\\
&\langle E| HO_{mn} |E\rangle=E \langle E| O_{mn}|E\rangle.
\label{HO=EO}
\end{align}
Here $E$ is the energy eigenvalue of the eigenstate $|E \rangle $.
(In the following, we omit $|E \rangle $.)
By substituting \eqref{observables} to these two equations, we obtain
\begin{align} 
	m^2 \langle O_{mn} \rangle +2m \langle O_{mn+1} \rangle+\frac{m\theta}{\pi} \langle O_{mn} \rangle
	+a\left[ \sum_{k=0}^{n-1} {}_n C_k \left( \langle O_{m+1k} \rangle + (-1)^{n-k} \langle O_{m-1k} \rangle  \right)   \right]
	=0,	
	\label{H-eq}
\end{align}
and
\begin{align} 
	&\frac{1}{2}
	\left[
		m^2 \langle O_{mn} \rangle +2m \langle O_{mn+1} \rangle + \langle O_{mn+2} \rangle+\frac{m\theta}{\pi} \langle O_{mn} \rangle+\frac{\theta}{\pi} \langle O_{mn+1} \rangle
		-a \left( \langle O_{m+1n} \rangle + \langle O_{m-1n} \rangle \right)   
	\right] \nonumber \\ 
	=& \left( E- \frac{\theta^2}{8\pi^2} -a \right) \langle O_{mn} \rangle.	
	\label{E-eq}
\end{align}
 The summation in \eqref{H-eq} appears when the operators in the equation are ordered into the forms \eqref{observables} through the commutator relation $[p,e^{imx}]=me^{imx}$.
By solving these equations\footnote{We can solve \eqref{H-eq} and \eqref{E-eq} analytically.
We can also solve them by using Mathematica. We use Mathematica because we can apply the same code to various potential case easily.}, 
we can describe all the observables  $\langle O_{mn} \rangle$ by the three variables: $\langle p \rangle$, $\langle e^{ i x} \rangle$ and $E$.
For example, from \eqref{H-eq} with $(m,n)=(1,0)$, we obtain
\begin{align} 
	 \left\langle e^{ i  x} p\right\rangle= - \left(\frac{1}{2} + \frac{\theta}{2\pi} \right) \left\langle e^{ i  x} \right\rangle.
\end{align}
By substituting them into the bootstrap matrix ${\mathcal M}$, we obtain,
\begin{align}
	{\mathcal M}=
	\begin{pmatrix}
	1 & \left\langle e^{  i x} \right\rangle  & \left\langle p \right\rangle   & \cdots \\
	\left\langle e^{  i x} \right\rangle  & 1 &  \left(\frac{1}{2} - \frac{\theta}{2\pi} \right) \left\langle e^{ i  x} \right\rangle &  \cdots\\
	\left\langle p	\right\rangle  &  \left(\frac{1}{2} - \frac{\theta}{2\pi} \right) \left\langle e^{ i  x} \right\rangle & 
	-2a+2E-\frac{\theta^2}{4 \pi^2} +2a\left\langle e^{  i x} \right\rangle 
	- \frac{\theta   }{\pi} \left\langle p \right\rangle & \cdots \\
	\vdots & \vdots  & \vdots  & \ddots
	\end{pmatrix}.
	\label{bootstrap2}
\end{align}
The idea of the bootstrap method is excluding the values of the variables $\langle p \rangle$, $\langle e^{ i x} \rangle$ and $E$ that do not satisfy the constraint ${\mathcal M} \succeq 0$.
Particularly, if the size of the bootstrap matrix ${\mathcal M}$ is sufficiently large, the allowed values might be very limited, and they might be almost identical to the values of the variables evaluated by energy eigenstates.

\begin{figure}
	\begin{tabular}{cc}
		\begin{minipage}{0.5\hsize}
			\begin{center}
				\includegraphics[scale=0.6]{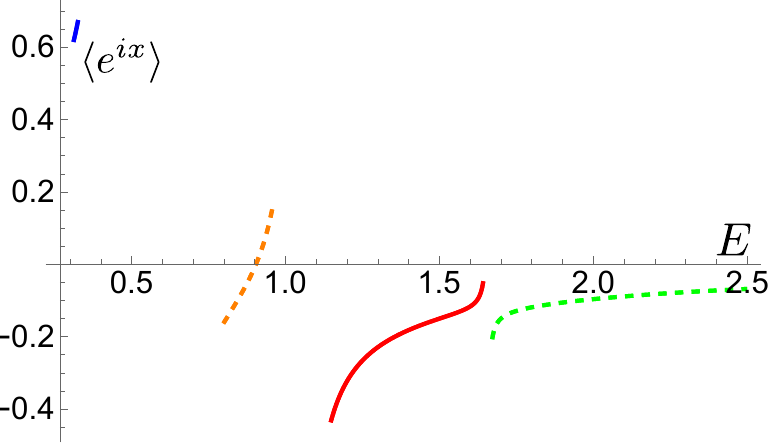}\\
				$E$ vs. $\langle e^{ix} \rangle$ at $\theta=0$
			\end{center}
		\end{minipage}		
		\begin{minipage}{0.5\hsize}
			\begin{center}
				\includegraphics[scale=0.6]{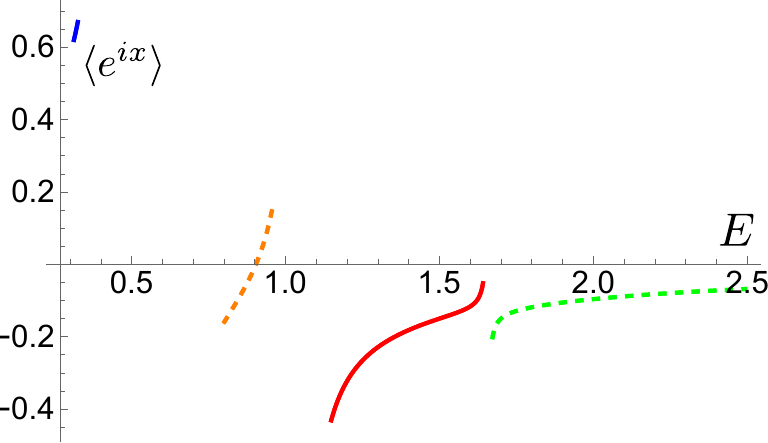}\\
				$E$ vs. $\langle e^{ix} \rangle$ at $\theta=\pi/2$
			\end{center}
		\end{minipage}
	\end{tabular}\\

	\caption{
		$E$ vs. $\langle e^{ix} \rangle$ through the bootstrap analysis.
		We take $a=1/2$ and investigate $\theta=0$ and $\pi/2$. 
		We fix $E$ and find the minimum (maximum) value of $\langle e^{ix} \rangle$ numerically,
		although the convergence is very good and we cannot distinguish the minimum and maximum in these plots.
		The curves are for the first four energy eigenstates considered in Fig.~\ref{Fig-a1} and Fig.~\ref{Fig-a1-obs}. 
        The line styles are the same as explained in Fig.~\ref{Fig-a1}.
		We find that the curves do not depend on $\theta$ almost and they are identical to the analytic results in Fig.~\ref{Fig-a1-obs}.
		}
	\label{Fig-a1-SDP-X}
\end{figure}

\begin{figure}
    \begin{tabular}{cc}
		\begin{minipage}{0.5\hsize}
			\begin{center}
				\includegraphics[scale=0.6]{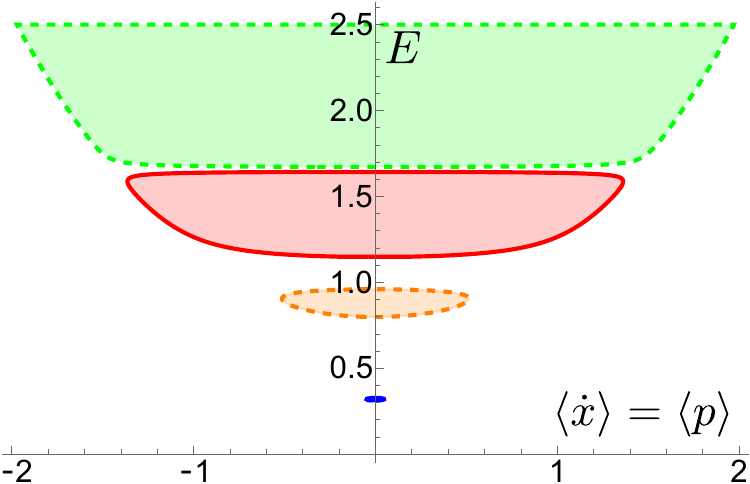}\\
				$\langle \dot{x} \rangle = \langle p \rangle $ vs. $E$ at $\theta=0$
			\end{center}
		\end{minipage}		
		\begin{minipage}{0.5\hsize}
			\begin{center}
				\includegraphics[scale=0.6]{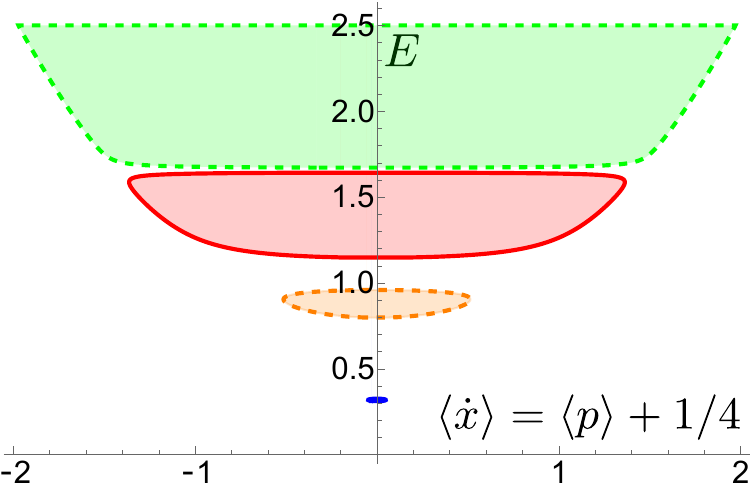}\\
				$\langle \dot{x} \rangle
				=\langle p \rangle +1/4$ vs. $E$ at $\theta=\pi/2$
 			\end{center}
		\end{minipage}
	\end{tabular}

	\caption{
        $\langle \dot{x} \rangle =\langle p \rangle +\theta/2\pi$ vs. $E$ for the first four eigenstates through the bootstrap analysis.
		We take $a=1/2$ and investigate $\theta=0$ (left panel) and $\pi/2$ (right panel). 
		We fix $E$ and find the minimum (maximum) value of  $\langle p \rangle$ numerically.
		The shaded regions are allowed region that satisfy ${\mathcal M} \succeq 0$.
		These regions do not depend on the values of $\theta$, and the boundaries of the regions are almost identical to the curves derived in the analytic calculation shown in Fig.~\ref{Fig-a1-obs} (left).
        	}
	\label{Fig-a1-SDP-P}
\end{figure}

More concretely, we assign some values to $a$, $\theta$ and $E$, and numerically find the possible minimum and maximum values of $\langle p \rangle$ (or $\langle e^{ i x} \rangle$) that satisfy the condition ${\mathcal M} \succeq 0$. This is called ``numerical bootstrap problem"\footnote{
	\label{ftnt-SDP}The bootstrap matrix ${\mathcal M}$ depends on $\langle p \rangle$ and $\langle e^{ i x} \rangle$ linearly  while on $E$ non-linearly.
    Thus, when we fix $E$, the problem finding the minimum (or maximum) of $\langle p \rangle$ (or $\langle e^{ i x} \rangle$) that satisfies ${\mathcal M} \succeq 0$ reduces to so called ``Semidefinite Programming Problem", and the numerical costs drastically decrease.
	In our numerical analysis, we use Mathematica package SemidefiniteOptimization, which is available in the version 12 or later.
	Note that the results of this package would depend on the option  ``Method" and we use ``MOSEK" in this work.
}. 
By repeating these computations by changing $E$, we obtain the results summarized in Fig.~\ref{Fig-a1-SDP-X} and \ref{Fig-a1-SDP-P} for $\theta=0$ and $\pi/2$ at $a=1/2$.

Firstly, we mention the convergence of our numerical results.
They converge sufficiently first.
Fig.~\ref{Fig-a1-SDP-X} and \ref{Fig-a1-SDP-P} are for $K_p=3$ and $K_x=4$, but, even if the size of the bootstrap matrix is increased, there is no change in the appearance of these figures.

Now, we discuss the details of our numerical results.
First, we consider the plot for $E$ vs. $\langle e^{ix} \rangle$ depicted in Fig.~\ref{Fig-a1-SDP-X}.
Surprisingly, the results do not depend on $\theta$ almost. 
$\theta=0$ and $\theta=\pi/2$ provide nearly same results.
Besides, the obtained curves are almost same to the analytic result shown in Fig.~\ref{Fig-a1-obs} (right).
However, there is a crucial difference.
The curve in Fig.~\ref{Fig-a1-obs} is the result for $0 \le \theta < 2\pi.$
Thus, if we fix $\theta$ to a single value, it corresponds to a single point on the curve.
(More precisely, the plot in Fig.~\ref{Fig-a1-obs} is for the four states, and four points appear at this $\theta$.)
On the other hand, the curves in Fig.~\ref{Fig-a1-SDP-X} are derived both at the single values of $\theta$.

This difference between the analytic result and the bootstrap method can be explained as follows.
In the former analysis, we have taken the gauge $\psi(x+2\pi)=\psi(x)$.
In the bootstrap analysis, however, we have not fixed the gauge.
Thus, even if $\theta$ in the Hamiltonian \eqref{H} is taken to be zero, if the periodicity of the state is $\theta/2\pi$, the situation is equivalent to $H(\theta)$ with the periodic state through the gauge transformation \eqref{gauge-tr}:
\begin{align} 
	H(0) \psi = E \psi, \quad \psi(x+2\pi) =e^{i\theta/2\pi} \psi(x)
 \quad \Longleftrightarrow  \quad
 H(\theta) \psi = E \psi, \quad \psi(x+2\pi) =\psi(x).
\end{align}
Hence, the bootstrap method derives the results for all possible values of $\theta$ in the Hamiltonian, even though we have taken $\theta=0$ and $\pi/2$ in Fig.~\ref{Fig-a1-SDP-X}.

If we wish to obtain the result for a fixed $\theta$ corresponding to Fig.~\ref{Fig-a1-obs}, we need to specify the periodicity of the states and fix the gauge.
However, we could not find a good way.
For example, we can read off the periodicity of the state by using the operator $e^{2\pi i p}$.
Indeed, if the wave function has a periodicity $\psi(x+2\pi)=e^{i \eta}\psi(x)$,
the corresponding state satisfies $\langle e^{2\pi i p} \rangle =e^{i \eta} $.
Thus, in order to fix the periodicity of the state, for example, $\psi(x+2\pi)=\psi(x)$, we should impose the additional constraint $\langle e^{2\pi i p} \rangle = 1$ in the bootstrap analysis.
However, the operator $e^{2\pi i p}$ commutes with all $O_{mn}$ in \eqref{observables}, and the condition $\langle e^{2\pi i p} \rangle = 1$ does not restrict the values of $O_{mn}$. 
Hence, this method might not work\footnote{After the authors submitted the first manuscript of this work to arxiv,
a related study \cite{Tchoumakov:2021mnh} appeared independently.
There, a possible way for determining  $\langle e^{2\pi i p} \rangle  $ was proposed, and it may overcome the issue of the gauge fixing.
}.

The difficulty of fixing the periodicity of the states causes a problem that we cannot determine the $\theta$-dependence of the observables such as $E(\theta)$ and $\langle e^{ix} \rangle (\theta)$, since we always obtain the result for all $\theta$.
This is a disadvantage of our bootstrap analysis.

Now, we move to discuss the details of the plot for $\langle \dot{x} \rangle$ vs. $E$ in Fig.~\ref{Fig-a1-SDP-P}.
Again, the results are almost independent of $\theta$\footnote{Note that if we plot $E$  vs. $\langle p \rangle$ rather than $\langle \dot{x} \rangle$ via the bootstrap method,
the results are almost same to Fig.~\ref{Fig-a1-SDP-P} but the center of the curves $\langle \dot{x} \rangle=0$ moves to $\langle p \rangle=-\theta/2\pi$.
Thus, the results slightly depend on $\theta$.
But $\langle p \rangle$ is not a physical observable and the gauge invariant quantity $\langle \dot{x} \rangle$ does not depend on $\theta$.
}, and the shapes of these curves agree with the analytic result in Fig.~\ref{Fig-a1-obs} (left) for all $\theta$.
This is similar to the $E$ vs. $\langle e^{ix} \rangle$ case.
However, there is one difference.
The inside regions of the curves satisfy the condition ${\mathcal M} \succeq 0$ and they are not excluded in the bootstrap analysis, while no solution appears there in the analytic result.

This subtle issue can be understood as follows.
Suppose that we fix $E$ and ask what is the possible value of $\langle \dot{x} \rangle$.
From Fig.~\ref{Fig-a1-obs} (left), we see that there are two possible values, say $\pm  \dot{x} (E)$.
Then, the general solution at given $E$ may be described as superposition of these two solutions.
If so, the possible values of the expectation value $\langle \dot{x} \rangle$ may be in the following range,
\begin{align}
-\dot{x} (E) \le \langle \dot{x} \rangle \le \dot{x} (E).
\label{dx-bootstrap}
\end{align}
This corresponds to the bootstrap results shown in Fig.~\ref{Fig-a1-SDP-P}.
However, this is not the end of the story.
Since the point on the curves in Fig.~\ref{Fig-a1-obs} is for the solution at a single $\theta$, the values of $\theta$ at $ \dot{x} (E)$ and  $- \dot{x} (E)$ are different.
Thus, we cannot superpose the solutions for $ \dot{x} (E)$ and  $- \dot{x} (E)$, and the regions inside the curves in Fig.~\ref{Fig-a1-SDP-P} should be excluded.

On the other hand, we have not fixed the periodicity of the states in the bootstrap analysis, and superposition of the $ \pm \dot{x} (E)$ solutions have not been excluded. This is consistent with the result in Fig.~\ref{Fig-a1-SDP-P}.
(Recall that different $\theta$ corresponds to the different periodicity through the gauge transformation \eqref{gauge-tr}.)
As we have argued, it is hard to fix the periodicity in the bootstrap method, and we cannot exclude the inside region.
However, it may not be a serious issue, since we can easily read off the correct correlations between $\langle \dot{x} \rangle$ and $E$: they appear at the boundaries of the regions in Fig.~\ref{Fig-a1-SDP-P}.

In this way, we have obtained the correlations among $\langle e^{ix} \rangle$, $\langle \dot{x} \rangle$ and $E$.
As we have mentioned, other operators $\langle O_{mn} \rangle$ are described by these three quantities, and we can easily obtain 
correlations among all the observables. 

Finally, we briefly mention the relation to the Bloch's theorem \cite{Tong:gauge}.
The Bloch's theorem states that the system $(-\infty < x < \infty)$ with a periodic potential would have a band structure.
Since we have not fixed the periodicity of the states, we may regard that our system is in $-\infty < x < \infty$ rather than $S^1$ \footnote{
The Hilbert space in $-\infty < x < \infty$ and that of on $S^1$ are different.	
We have implicitly assumed that the system is in $-\infty < x < \infty$ when we derive \eqref{dx-bootstrap}.}.
Then, the energy curves in Fig.~\ref{Fig-a1-SDP-X} and \ref{Fig-a1-SDP-P} naturally correspond to the band structure.

\subsection{$\theta=0$ and $\pi$}

So far, we have seen that the bootstrap method can derive the correlations among the observables, but it cannot derive the $\theta$ dependence of them.
Here, we argue that actually $\theta=0$ and $\pi$ are special, and we can read off the values of the observables there.

The Hamiltonian \eqref{H} is invariant under the parity symmetry: $(x,p,\theta) \to (-x,-p,-\theta) $.
Thus, the eigenstates can be taken parity even or odd, and the expectation value of $p$ at a given $\theta$ satisfies
\begin{align}
	p (-\theta) = -  p (\theta),
   \end{align}
where we have omitted the bra-ket symbols and taken the periodic gauge.
Besides, through the gauge symmetry \eqref{gauge-tr} and \eqref{gauge-tr-p}, we have
\begin{align}
	p (\theta-2 \pi) =   p (\theta)+1.
\label{gauge-tr-p-2}
\end{align}
By combining these two equations with $\theta=0$ and $\pi$, we obtain
\begin{align}
	p (0) =  0, \qquad p (\pi) =  -\frac{1}{2}.
\end{align}
Actually, the analytic result is consistent with this relation as shown in Fig.~\ref{Fig-a1} (center).
Note that this result implies that the velocity $\langle \dot{x} \rangle = \langle p \rangle + \theta/2\pi $ always becomes zero at $\theta=0$ and $\pi$.

Now, we consider the application of this result to the bootstrap analysis to determine the $\theta$ dependence at $\theta=0$ and $\pi$.
In Fig.~\ref{Fig-a1-SDP-P}, we have derived the curves $E(\langle \dot{x} \rangle)$, and we can read off $E$ at $\langle \dot{x} \rangle = 0 $.
Since $\langle \dot{x} \rangle$ becomes zero at  $\theta=0$ and $\pi$, they may correspond to $E(\theta)=E(0)$ or $E(\pi)$.
Here, we can determine whether these values are for $\theta=0$ or $\pi$ by using the result at the $a=0$ case discussed in Appendix \ref{app-a=0}.

Our model \eqref{H} at $a>0$ can be regarded as a deformation of the $a=0$ case, and we can easily see that the curves in Fig.~\ref{Fig-a1-SDP-P} at $a >0 $ merge to the ones shown in Fig.~\ref{Fig-a0} (right) as $a \to 0$.
Besides, Fig.~\ref{Fig-a0} tells us the values of $E$, $p$ and $\theta$ at the point $\langle \dot{x} \rangle = 0 $ for $a=0$ as
\begin{align} 
(E,p ,\theta)=(0,0,0),~(1/8, -1/2,\pi),~(1/2,0,0),~ (9/8, -1/2,\pi),~ (2, 0,0),~\cdots.
\label{E-E-p-theta-a=0}
\end{align}
Since these points would continue to $E(\langle \dot{x} \rangle)=E(0)$ at $a>0$, we can fix the values of $\theta$ for $E(0)$ from them.
For example, for the first eigenstate at $a>0$, there are two $E(\langle \dot{x} \rangle)=E(0)$ as we see in Fig.~\ref{Fig-a1-SDP-P}, and we can determine that the lower energy is for $\theta=0$ and the higher one is for $\theta=\pi$ through \eqref{E-E-p-theta-a=0}.
Thus, we obtain $E(\theta=0) \simeq 0.31$ and $E(\pi) \simeq 0.33$ at $a=1/2$ from Fig.~\ref{Fig-a1-SDP-P}.

Similarly, at the second eigenstate, the lower energy is for $\theta=\pi$ and the higher one is for $\theta=0$.
Generally, at the $(2n+1)$-th eigenstate, the lower one is for $\theta=0$ and the higher one is for $\theta=\pi$, and it is opposite at the $2n$-th eigenstate.
They agree with the analytic result shown in Fig.~\ref{Fig-a1} (left).
In this way, we can determine $E(\theta=0)$ and $E(\pi)$\footnote{Our derivation relys on the information of the $a=0$ result.
If we consider different systems with $\theta$-terms, we may need to use alternative inputs.
For example, see \cite{Witten:1998uka} for large-$N$ gauge theories.
}.
Once we obtain the energies at $\theta=0$ and $\pi$, all other observables are determined through the correlations discussed above.

\section{Discussions}
\label{sec-discussion}

In this article, we studied the $\theta$ dependence in the model \eqref{S}, and found that the correct correlations among the observables can be derived.
However, we failed to obtain the $\theta$ dependence of the observables except at $\theta=0$ and $\pi$ because of the gauge fixing problem.
Hence, the bootstrap method tells us, for example, that $E(\theta)$ for the $n$-th energy eigenstates is in the range shown in Fig.~\ref{Fig-a1-SDP-X}, and if one value of energy is given in this range, the values of the observables $\langle O_{mn} \rangle $ are determined.
Thus, the bootstrap method provides us useful information but not as much as the analytic results derived in Sec.~\ref{sec-analytic}.

Our study reveals two interesting properties of the bootstrap method.
The first one is that the method may work even if the system suffers sign problems, although it may fail in evaluating some quantities.
We presume that, as far as the condition  $ \langle O^\dagger O \rangle \ge 0$ is satisfied for suitable observables $O$,
the bootstrap method may work somehow\footnote{For real time evolution, although the condition $ \langle O^\dagger O \rangle \ge 0$ is satisfied, \eqref{HO=EO} is not satisfied and \eqref{HO=0} is replaced by the Heisenberg equation.
Thus, the constraints are much weakened.
Besides, the Heisenberg equations are differential equations and they may be incompatible with the inequality constraint ${\mathcal M} \succeq 0$.
A similar issue occurs even in thermal equilibrium states.
We will report this problem soon \cite{AMY}.
}.

Another interesting property is that the bootstrap method may illuminate hidden natures of systems.
Suppose that we did not know the existence of the $\theta$-parameter in the system \eqref{S} and studied \eqref{H} with $\theta=0$.
Even in this case, the bootstrap method reproduces the results that take the $\theta$ dependence into account.
Hence, by applying the bootstrap method to various systems, some unexpected new phenomena might be found.

One important challenge is overcoming the gauge fixing problem that prevents us from deriving the $\theta$ dependence of the observables.
Another challenge is application of the bootstrap method to higher dimensional quantum field theories.
If it is achieved, we can apply the bootstrap method to QCD and tackle the $\theta$-term problem there.

Besides, there are several interesting large-$N$ matrix models in zero and one dimensions, which suffer sign problems.
(For example, the Lorentzian IKKT matrix model \cite{Ishibashi:1996xs, Kim:2011cr} and the BFSS matrix theory \cite{Banks:1996vh, Anagnostopoulos:2007fw}.)
Thus, it must be valuable to test the bootstrap method in these models.

\paragraph{Acknowledgements}
The authors would like to thank Takehiro Azuma, Masaru Hongo and Asato Tsuchiya for valuable discussions and comments.
A part of numerical computation in this work was carried out at the Yukawa Institute Computer Facility.
The work of T.~M. is supported in part by Grant-in-Aid for Scientific Research C (No. 20K03946) from JSPS.

\appendix

\section{Analytic result at $a=0$}
\label{app-a=0}

In this appendix, we summarize the analytic results at $a=0$ in \eqref{H}, in which we can derive the solutions easily.
The results at $a>0$ can be regarded as deformations of $a=0$ and the analysis at $a=0$ may help us to understand the properties of the system at $a>0$.

\begin{figure}
	\begin{tabular}{ccc}
		\begin{minipage}{0.33\hsize}
			\begin{center}
				\includegraphics[scale=0.4]{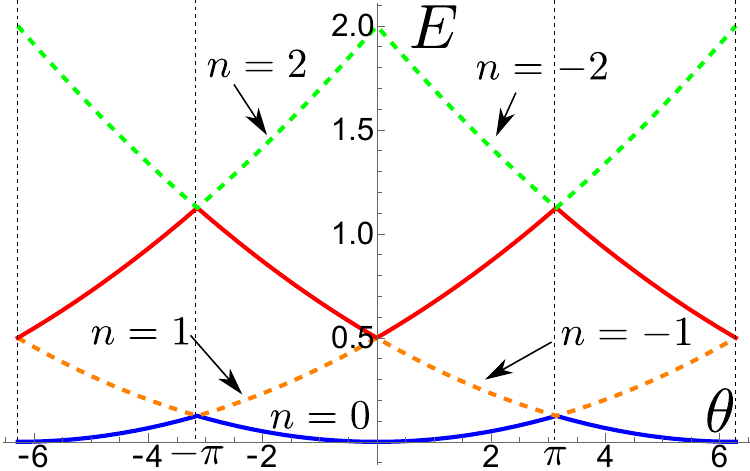}\\
				 $\theta$ vs. $E$
			\end{center}
		\end{minipage}
		\begin{minipage}{0.33\hsize}
			\begin{center}
				\includegraphics[scale=0.4]{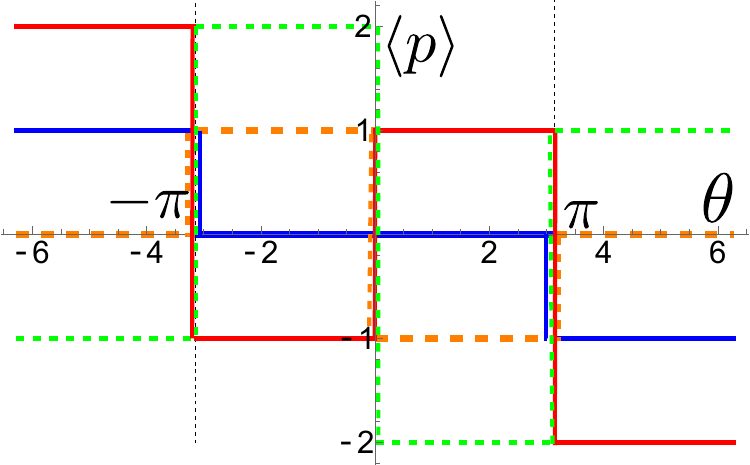}\\
				$\theta$ vs. $\langle p \rangle$
			\end{center}
		\end{minipage}
		\begin{minipage}{0.33\hsize}
			\begin{center}
				\includegraphics[scale=0.4]{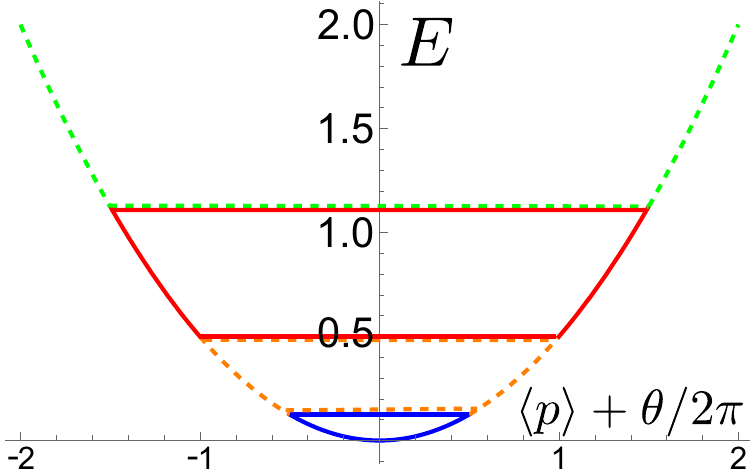}\\
				$\langle \dot{x} \rangle=\langle p \rangle +\theta/2\pi $ vs. $E$
			\end{center}
		\end{minipage}
	\end{tabular}
	\caption{
	Analytic results at $a=0$.
	We plot $E$ and $\langle p \rangle$  for the first four eigenstates.
	The line styles correspond to those of Fig.~\ref{Fig-a1}.
	Although the graphs $\theta$ vs. $E$ and $\theta$ vs. $\langle p \rangle$ are involved,
	the combined result $\langle\dot{x}  \rangle $ vs. $E$ becomes simpler.
		}
	\label{Fig-a0}
\end{figure}

Under the gauge fixing $\psi(x+2\pi)=\psi(x)$, we obtain the energy eigenfunction and its energy eigenvalue,
\begin{align}
\phi_n(x):=\frac{1}{\sqrt{2\pi}} e^{inx} ,\quad E_n:= \frac{1}{2} \left(
n+\frac{\theta}{2\pi}
\right)^2.
\label{a=0}
\end{align}
Interestingly, the energy levels change depending on the values of $\theta$.
See Fig.~\ref{Fig-a0} (left).
At $\theta=0$, $n=0$ is the ground state, and the first excited state is $n=\pm1$ and the degeneracy occurs. Higher excitations are given by $n= \pm 2, \pm 3 , \cdots$.
For $0 < \theta < \pi$, the degeneracies at the excited states disappear. The ground state is $n=0$, the first excited state is $n=-1$, and the second one is $n=1$.
At $\theta=\pi$, the ground state is $n=0$ and $n=-1$, and they are degenerate.
For $\pi < \theta < 2\pi$, the ground state is $n=-1$.
In this way, although the Hamiltonian \eqref{H} at $a=0$ is simple, the spectra show complicated $\theta$ dependence.

However, if we use the velocity $\langle	\dot x  \rangle =\langle p \rangle + \theta/ 2\pi$ \eqref{velocity}, the complicated spectrums can be simplified.
Here, the expectation value of the momentum $\langle p \rangle$ is easily computed, since $\phi_n$ in \eqref{a=0} is the eigenfunction of $p$ with the eigenvalue $n$. 
Then, the $\theta$ dependence of $\langle p \rangle $ is plotted as in Fig.~\ref{Fig-a0} (center).
Note that, due to the degeneracies at $\theta=k \pi$, ($k \in {\mathbf Z}$), $\langle p \rangle $ becomes multiple values there.
Although this $\theta$ dependence is involved, by combining \eqref{a=0} and \eqref{velocity}, we obtain a simplified expression with respect to  $\langle	\dot x  \rangle$,
\begin{align}
	E=\frac{1}{2} 	\left( \langle p \rangle  + \frac{\theta}{2\pi}   \right)^2
	=\frac{1}{2} \langle	\dot x  \rangle^2 .
	\label{E-a=0}
\end{align}
See Fig.\ref{Fig-a0} (right).
Again, due to the degeneracies at $\theta= k \pi$, multiple values appear at $E=k^2/8$.

If we compare the results at $a=0$ (Fig.~\ref{Fig-a0}) and the ones at $a>0$ (Fig.~\ref{Fig-a1} and \ref{Fig-a1-obs}), we find that the latter can be regarded as a deformation of the former.

\paragraph{Bootstrap analysis}
Bootstrap analysis at $a=0$ is special, since momentum is conserved and the energy eigenstate can be a momentum eigenstate.
Thus, $p$ can be treated as a c-number, and we obtain \eqref{E-a=0} directly.
Besides, we can fix the gauge because $\langle e^{2\pi i p}  \rangle= e^{2\pi i \langle p \rangle}  $ and the periodicity of the states is controlled by the value of $p$.
If we wish to take the gauge $\psi(x+2\pi)=\psi(x)$, $p$ is restricted to integers,
and we reach the energy \eqref{a=0}.
Thus, without using a bootstrap, we obtain the solution merely through the symmetries.

On the other hand, it is possible to perform the standard bootstrap analysis, if we do not impose the condition that the states are eigenstates of momentum.
Then, we obtain similar results to the ones in Sec.~\ref{sec-bootstrap} at $a>0$.

\section{The Numerical Bootstrap in Mathematica}
\label{app-bootstrap}

In this appendix, we briefly explain how to implement the numerical bootstrap method by using Mathematica\footnote{ A sample code of our numerical bootstrap analysis is available at \url{https://www2.yukawa.kyoto-u.ac.jp/~takeshi.morita/}}.

Firstly, we need to treat operators such as $e^{ix}$ and $p$, which do not commute each other.
The point is that, through the commutation relations, any operator can reduce to a sum of 
the ordered operators $e^{imx}p^n$ (not $p^ne^{imx}$ in the following explanation).
Thus, to express operators in Mathematica, we define the object
\begin{align}
	\text{op}[m,n]:=e^{imx}p^n.
\end{align}
We also define a function for computing the product of two operators $\text{op}[k,l]$ and $\text{op}[m,n]$ as
\begin{align}
	\text{prod}[\text{op}[k,l],\text{op}[m,n]]:=&e^{ikx}p^l e^{imx}p^n
	= \sum_{r=0}^l m^r {}_l C_r  e^{i(k+m)x}p^{l+n-r} \nonumber \\
	=& \sum_{r=0}^l m^r {}_l C_r  \text{ op}[k+m,l+n-r].
	\label{math-prod}
\end{align}
Here ${}_l C_r:=l!/(l-r)!/r!$ and we have used the commutation relation $[p,e^{imx}]=me^{imx}$.
Then, by using this function, we can easily compute various equations involving operators in Mathematica.
For example, if we want to calculate $\text{eq}1 \times \text{eq}2$, where $\text{eq}1:=5p^2+e^{ix} = 5~\text{op}[0,2]+\text{op}[1,0]$ and $\text{eq}2:=3e^{i2x}-p=3~\text{op}[2,0]-\text{op}[0,1]$,
we can do it through the following code:
\begin{align}
&	\text{var1}=\text{Variables}[\text{eq1}]; \nonumber \\
&	\text{coeff1}=\text{Coefficient}[\text{eq1},\text{var1}]; \nonumber \\
&	\text{var2}=\text{Variables}[\text{eq2}]; \nonumber \\
&	\text{coeff2}=\text{Coefficient}[\text{eq2},\text{var2}]; \nonumber \\
&	\text{Flatten}[\text{Outer}[\text{Times},\text{coeff1},\text{coeff2}]].\text{Flatten}[\text{Outer}[\text{prod},\text{var1},\text{var2}]].
\label{product-eqs}
	\end{align}
	Here we obtain $\text{var1}=\{\text{op}(0,2),\text{op}(1,0)\}$ and $\text{coeff1}=\{5,1\}$, and will obtain the correct product of the operators.
It is convenient to define this procedure as a single function.
Now, we are ready to start the numerical bootstrap program.
The procedure is as follows:
\begin{enumerate}
	\item Constructing the bootstrap matrix ${\mathcal M}$ \eqref{bootstrap}. 
	We define the seed operators $\{ \text{op}[m,n] \}$ for \eqref{operators-bootstrap} and their Hermite conjugate 
	$\{ (\text{op}[m,n])^\dagger \}=\{ \text{prod}[ \text{op}[0,n], \text{op}[-m,0]] \}$.
	Each component of the bootstrap matrix is the product of these operators, which can be calculated through \eqref{product-eqs}. 
	\item Solving the constraint equations \eqref{HO=0} and \eqref{HO=EO}. 
	We can do it analytically by hand, and represent all the operators by using $\text{op}[0,0]$, $\text{op}[0,1]$ and $\text{op}[1,0]$.
	However, if we want to consider more general Hamiltonian, it might be better to compute $[ H, O_{mn} ] $ and $ HO_{mn}$, and  solve them by using Mathematica.
	\item Performing the optimization.
	We substitute the solution of the constraint equation \eqref{HO=0} and \eqref{HO=EO} into the bootstrap matrix ${\mathcal M}$.
	Then, we fix a value of $E$ and find the minimum value of the quantity that we are interested in, for example $\langle e^{ix} \rangle=\text{op}[1,0]$, under the constraint ${\mathcal M} \succeq 0$ by using the package ``SemidefiniteOptimization".
	``FindMinimum" is also available but usually ``SemidefiniteOptimization" is much efficient. See footnote \ref{ftnt-SDP} also. 
\end{enumerate}

{\normalsize 
\bibliographystyle{unsrt}
 \bibliography{bBFSS} }

\end{document}